\documentclass[aps,pra,reprint,superscriptaddress,nofootinbib]{revtex4-2}

\usepackage{amsmath,amssymb,amsfonts,mathtools,bm}
\usepackage{braket}
\usepackage{graphicx}
\usepackage{hyperref}

\usepackage[dvipsnames]{xcolor}

\definecolor{DSPink}{RGB}{204,51,119}
\DeclareRobustCommand{\DS}[1]{{#1}}
\newcommand{\tc}[1]{{#1}}
\long\def\tc#1{{#1}}

\newcommand{\GJ}[1]{{{#1}}}

\begin{document}

\title{Single-realization Pauli twirling for suppressing coherent error accumulation in structured quantum circuits}

\author{Dihang Sun}
\affiliation{Department of Physics, National University of Singapore, Singapore 117551, Singapore}
\affiliation{Centre for Quantum Technologies, National University of Singapore, Singapore 117543, Singapore}

\author{Tianqi Chen}
\affiliation{Institute of Advanced Intelligence and Computing (IAIC), Agency for Science, Technology and Research (A*STAR), 1 Fusionopolis Way, No.~16-16 Connexis, Singapore 138632, Singapore}
\affiliation{Bioinformatics Institute, Agency for Science, Technology and Research (A*STAR), 30 Biopolis Street, No.~07-01 Matrix, Singapore 138671, Singapore}

\author{Jin Ming Koh}
\affiliation{Department of Physics, Harvard University, Cambridge, Massachusetts 02138, USA}
\affiliation{Harvard--MIT Center for Ultracold Atoms, Cambridge, Massachusetts 02138, USA}
\affiliation{The NSF AI Institute for Artificial Intelligence and Fundamental Interactions}

\author{Nannan Ma}
\affiliation{Department of Physics, National University of Singapore, Singapore 117551, Singapore}
\affiliation{Centre for Quantum Technologies, National University of Singapore, Singapore 117543, Singapore}

\author{Jiangbin Gong}
\email{phygj@nus.edu.sg}
\affiliation{Department of Physics, National University of Singapore, Singapore 117551, Singapore}
\affiliation{Centre for Quantum Technologies, National University of Singapore, Singapore 117543, Singapore}

\begin{abstract}
Pauli twirling is conventionally understood as an ensemble-level noise-tailoring procedure which suppresses coherent components of the error channel by averaging over randomized compilations. This picture does not by itself determine whether a single realization of Pauli-dressed circuit can outperform its bare counterpart, nor does it predict when such an advantage should occur. In this work, we \GJ{present single-realization based framework for the standard Pauli dressing of two-qubit Clifford gates in structured quantum circuits.  This is afforded by the key insight that for weak coherent errors, the leading gate infidelity is governed by the variance of some accumulated error generator}. Through  \GJ{both simulations and theoretical considerations}, we verify that coherent errors accumulated at different entangling gates can produce $\Theta(L^2)$ variance growth in a bare circuit, whereas independent Pauli dressing eliminates the cross-layer contribution on average and yields an $O(L)$ variance bound at any fixed confidence level for a \GJ{single} randomly sampled realization. We then develop a covariance diagnostic that predicts when this mechanism is effective in error mitigation. We find a strong Pauli twirling advantage for coherent errors whose transported generators have large off-diagonal covariance, and little advantage when the covariance matrix is nearly diagonal. Hardware experiments on IBM superconducting processors demonstrate
fidelity improvements for GHZ-state preparation and reduced observable errors in Trotterized transverse-field Ising dynamics. These results indicate that single realizations of Pauli twirling without ensemble averaging is a useful coherent-error suppression technique for structured circuits containing repeated nontrivial two-qubit operations. 
\end{abstract}

\maketitle
\section{Introduction}

\tc{
Two-qubit entangling gates are a dominant source of error in present-day quantum processors. Their error rates are typically much higher than those of single-qubit operations~\cite{arute2019quantum,google2025qec}. Such entangling gates play an important role in a broad
range of quantum-state preparation and dynamical
simulation tasks on superconducting quantum processors
\cite{mooney2021generation,kim2023evidence,
ding2026manipulation}. In structured state-preparation and simulation circuits containing many such gates, coherent noise, such as small over-rotations or unwanted phases, can be transported into similar directions by the ideal circuit evolution and accumulate constructively \cite{wallman2015estimating,sanders2016bounding,kueng2016comparing,gutierrez2016errors}. This accumulation poses a particular challenge in the noisy intermediate-scale quantum (NISQ) regime, where large-scale fault-tolerant quantum computation \GJ{remains beyond the capabilities of present-day processors} \cite{preskill2018quantum,cai2023quantum}. Therefore, developing practical techniques to suppress coherent error accumulation while introducing minimal additional control errors remains an important goal for near-term quantum computing.

To improve the reliability of quantum computations, various quantum error mitigation (QEM) and noise-tailoring techniques have been developed \cite{cai2023quantum}. Among these approaches, Pauli twirling (PT) provides an attractive strategy for suppressing the detrimental effects of coherent noise. Pauli twirling is a noise-tailoring protocol which averages a noise channel over Pauli conjugations, yielding a stochastic Pauli channel \cite{emerson2007symmetrized,dankert2009exact}. Randomized compiling (RC) incorporates such noise tailoring into quantum circuits by inserting random single-qubit gates and compensating corrections, so that every randomized realization implements the same ideal computation \cite{wallman2016noise}. Under appropriate noise assumptions, averaging over these realizations converts the noise into a probabilistic mixture of Pauli errors, \GJ{thereby suppressing the detrimental effects of coherent error accumulation. This type of noise-tailoring technique} has been demonstrated through Pauli-frame randomization and randomized compiling on superconducting processors \cite{ware2021experimental,hashim2021randomized,hashim2023benchmarking}. For Clifford entangling gates, \GJ{the randomized compiling procedure} can be implemented by preceding each gate with an independently sampled two-qubit Pauli operator and following it with a matched Pauli correction \cite{wallman2016noise,hashim2021randomized}. Both the randomization and the matched Pauli corrections use only single-qubit operations, whose higher fidelity makes this construction well suited to suppressing the effects of entangling-gate errors with little additional control error.

Randomization has also been employed in other error-suppression protocols. For example, random dynamical decoupling uses random control pulses to preserve quantum states against unwanted Hamiltonian evolution and system-environment coupling \cite{viola2005random}. The Pauli-random-error-correction method adopts random Pauli operations to suppress errors caused by weak, unwanted interactions between qubits that persist between gate operations \cite{kern2005quantum}. Random-walk-like error accumulation has also been analyzed by the robust error accumulation suppression (REAS) protocol \cite{odake2025robust}, which combines randomization with error calibration and rotation-angle correction. Despite these advances, an important distinction remains between ensemble-averaged noise tailoring and the performance of an individual randomized circuit. Although previous studies have shown that randomized compiling can suppress coherent error accumulation \cite{winick2022concepts,hashim2023benchmarking}, the conventional description of PT primarily concerns the ensemble-averaged noise channel. 
This description does not, by itself, determine whether an individual Pauli-dressed circuit can outperform its bare counterpart, or under what conditions such an advantage can be expected. In particular, a state- and circuit-dependent understanding of how coherent errors generated at different entangling gates propagate and interfere remains valuable for predicting the effectiveness of PT in structured quantum circuits.

Motivated by these questions, in this work we investigate the suppression of coherent error accumulation by single realizations of standard Pauli dressing in structured quantum circuits containing repeated two-qubit Clifford gates. \GJ{We find that single realizations of PT can already substantially outperform their bare counterparts, even without using other error mitigation protocols. This effect is particularly relevant for circuits containing many noisy entangling gates, where errors originating from different gates may accumulate coherently. While a single realization of Pauli dressing does not generally eliminate the errors introduced by individual gates, randomization across multiple entangling layers can substantially suppress their accumulated effects. Understanding how a single realization of PT can mitigate errors calls for an analysis of errors generated at different entangling gates, their propagation, and their interference.} To this end, we develop a circuit-resolved perturbative theory for standard two-qubit Clifford Pauli dressing, which requires neither error calibration nor rotation-angle correction.

\GJ{The main finding at the single-realization level is as follows. Pauli dressing randomizes the signs of the Pauli components of the coherent error generators. In the weak-error regime, the leading state infidelity is found to be governed by the variance of the accumulated error generator associated with the input state. In the bare circuit, error contributions from different entangling layers can reinforce one another, producing quadratic growth of this variance, whereas independent Pauli dressing eliminates the cross-layer contributions upon averaging over the random Pauli choices. For bounded layerwise coherent error strengths, the resulting variance grows at most linearly on average, with a corresponding linear bound holding at any fixed confidence level for a randomly sampled realization. This qualitative insight also enables us to identify when constructive interference between errors at different gates can or cannot be effectively suppressed.} Specifically, a covariance matrix defined via the error generators becomes one important object: its diagonal entries quantify the intrinsic contributions of individual entangling layers, whereas its off-diagonal entries measure the signed overlap between the error directions generated at different layers. A substantial net positive contribution from the off-diagonal covariance identifies coherent reinforcement and therefore predicts a strong potential PT advantage at the single-realization level, whereas an approximately diagonal matrix indicates that little cross-layer coherent error accumulation is available to suppress. We illustrate the diagnostic in a fan-like GHZ-state-preparation circuit, using different types of generators as examples of constructive and nearly vanishing cross-layer covariance. These results show that the covariance matrix structure provides a state- and circuit-dependent criterion for predicting the effectiveness of single-realization Pauli twirling.

We further test our theoretical understanding and the effectiveness of single-realization PT on IBM superconducting quantum processors. These benchmarks connect our circuit-dependent analysis to applications of randomized compiling in quantum algorithms and many-body dynamics \cite{hashim2021randomized,gu2023noise,kurita2023synergetic,perrin2024mitigating,perrin2025dynamic,ishii2025implementation}. We find that for multiqubit GHZ-state preparation, PT systematically improves the mean measured fidelity over a broad range of system sizes. In a Trotterized transverse-field Ising simulation, PT reduces the deviations of the measured imbalance and local magnetization profiles from the noiseless Trotter evolution. An additional benchmark using a decomposed quantum Fourier transform illustrates the input dependence of the PT improvement. These results demonstrate that gate-protected randomization can suppress significant coherent two-qubit error accumulation across distinct circuit architectures and computational tasks.

This paper is organized as follows. Section~\ref{sec:theory} develops the two-qubit-gate PT construction for a general pure-state preparation circuit and derives its ensemble-averaged and single-realization properties. Section~\ref{sec:numerics} first verifies the predicted variance scaling in mirror circuits, then applies the theory to GHZ-state preparation and establishes the connected covariance matrix as a diagnostic of suppressible coherent error. Section~\ref{sec:hardware} presents the hardware results for GHZ-state preparation and Trotterized Hamiltonian simulation. The main findings are summarized in Sec.~\ref{sec:conclusion}.
}

\section{Theory: two-qubit-gate Pauli twirling}
\label{sec:theory}


\subsection{Pauli Twirling construction for a general circuit}

We consider a quantum circuit whose ideal output can be written as
\begin{equation}
\ket{\psi_{\rm tar}}
=
U_{\rm ideal}\ket{\psi_0},
\label{eq:general_target_state}
\end{equation}
where $\ket{\psi_0}$ is the input state. The circuit contains a sequence of selected two-qubit Clifford entangling
gates \(U_1,U_2,\ldots,U_L\), interleaved with deterministic layers \(B_\ell\) that
are not twirled. We write
\begin{equation}
U_{\rm ideal}
=
B_LU_LB_{L-1}U_{L-1}\cdots B_1U_1B_0,
\label{eq:general_circuit}
\end{equation}
where the layers \(B_\ell\) are either identities or contain only single-qubit gates. In particular, a non-Clifford two-qubit block may first be decomposed into Clifford
entangling gates and single-qubit rotations, after which the protocol
is applied only to the Clifford entangling gates.

For each selected two-qubit Clifford gate \(U_\ell\), we independently
sample a two-qubit Pauli operator
\begin{equation}
P_\ell
=
P_\ell^{(1)}\otimes P_\ell^{(2)}
\in\mathcal P_2,
\qquad
\mathcal P_2
=
\{I,X,Y,Z\}^{\otimes 2},
\label{eq:general_pre_pauli}
\end{equation}
uniformly at random. We then define the matched post-gate correction
\begin{equation}
Q_\ell
=
U_\ell P_\ell U_\ell^\dagger .
\label{eq:general_Q_def}
\end{equation}
Since \(U_\ell\) is a Clifford gate, conjugation by \(U_\ell\) maps
the Pauli group to itself \cite{gottesman1997stabilizer}. Hence \(Q_\ell\) is again a two-qubit Pauli
operator and can be implemented as a product of single-qubit Pauli
gates on the two qubits involved in \(U_\ell\).

The dressed realization of the ideal gate is 
\begin{equation}
V_\ell(P_\ell)
=
Q_\ell U_\ell P_\ell .
\label{eq:general_dressed_gate}
\end{equation}
Using Eq.~\eqref{eq:general_Q_def}, we obtain
\begin{equation}
V_\ell(P_\ell) =
\left(
U_\ell P_\ell U_\ell^\dagger
\right)
U_\ell P_\ell
\nonumber =
U_\ell P_\ell^2
=
U_\ell ,
\label{eq:general_gate_protection}
\end{equation}
up to an irrelevant global phase. Thus every Pauli-dressed
realization implements exactly the same ideal two-qubit gate.

Replacing all selected gates in Eq.~\eqref{eq:general_circuit} by their dressed realizations, we have
\begin{equation}
V_{\rm id}(\vec P)\ket{\psi_0}
=
\ket{\psi_{\rm tar}}
\end{equation}
for every realization in the absence of noise. The randomization
therefore changes only the physical implementation and its associated error process, while preserving the intended unitary state or the target state. 

\subsection{\tc{Conventional Pauli twirling: the ensemble-averaged channel}}

We first analyze the protocol at the channel level. For clarity, we
focus on the error associated with the selected two-qubit gates; errors
in the fixed layers \(B_\ell\) may be included separately but are not
twirled by the present construction. 
Let the noisy implementation of
the \(\ell\)-th selected entangling gate be
\begin{equation}
\widetilde{\mathcal U}_\ell
=
\mathcal U_\ell\circ\mathcal E_\ell,
\label{eq:general_noisy_channel}
\end{equation}
where
\begin{equation}
\mathcal U_\ell(\rho)
=
U_\ell\rho U_\ell^\dagger
\end{equation}
is the ideal channel and \(\mathcal E_\ell\) is a gate-local error
channel. The noisy dressed realization is
\begin{equation}
\widetilde{\mathcal V}_\ell^{(P_\ell)}
=
\mathcal Q_\ell
\circ
\widetilde{\mathcal U}_\ell
\circ
\mathcal P_\ell,
\label{eq:general_noisy_dressed_channel_start}
\end{equation}
where
\begin{equation}
\mathcal P_\ell(\rho)
=
P_\ell\rho P_\ell^\dagger,
\qquad
\mathcal Q_\ell(\rho)
=
Q_\ell\rho Q_\ell^\dagger .
\end{equation}
From Eq.~\eqref{eq:general_Q_def}, one has
\begin{equation}
Q_\ell U_\ell
=
U_\ell P_\ell .
\end{equation}
It follows that
\begin{align}
\widetilde{\mathcal V}_\ell^{(P_\ell)}
&=
\mathcal Q_\ell
\circ
\mathcal U_\ell
\circ
\mathcal E_\ell
\circ
\mathcal P_\ell
\nonumber\\
&=
\mathcal U_\ell
\circ
\left(
\mathcal P_\ell
\circ
\mathcal E_\ell
\circ
\mathcal P_\ell
\right).
\label{eq:general_dressed_channel}
\end{align}
Thus the ideal logical gate remains unchanged, while the error channel
is conjugated by the randomly sampled Pauli operator.

Averaging uniformly over \(P_\ell\in\mathcal P_2\) yields
\begin{equation}
\mathbb E_{P_\ell}
\widetilde{\mathcal V}_\ell^{(P_\ell)}
=
\mathcal U_\ell
\circ
\mathcal T_{\mathcal P_2}(\mathcal E_\ell),
\label{eq:general_averaged_channel}
\end{equation}
where
\begin{equation}
\mathcal T_{\mathcal P_2}(\mathcal E)
=
\frac{1}{16}
\sum_{P\in\mathcal P_2}
\mathcal P
\circ
\mathcal E
\circ
\mathcal P
\label{eq:general_pauli_twirl}
\end{equation}
is the two-qubit Pauli twirl of \(\mathcal E\). With the convention in
Eq.~\eqref{eq:general_noisy_channel}, the ensemble-averaged dressed
gate is therefore the same ideal gate preceded by the Pauli-twirled
error channel.

To make the effect of the twirl explicit, let \(R(\mathcal E)\) denote
the Pauli-transfer matrix of \(\mathcal E\). In the two-qubit Pauli
basis \(\{P_\mu\}\), its entries may be written as
\begin{equation}
R_{\mu\nu}(\mathcal E)
=
\frac{1}{4}
{\rm Tr}
\left[
P_\mu\mathcal E(P_\nu)
\right].
\label{eq:general_ptm_definition}
\end{equation}
Pauli twirling removes all off-diagonal entries of this matrix \cite{emerson2007symmetrized,dankert2009exact,wallman2016noise},
\begin{equation}
R\!\left[
\mathcal T_{\mathcal P_2}(\mathcal E)
\right]
=
{\rm diag}\!\left[
R(\mathcal E)
\right].
\label{eq:general_ptm_diagonal}
\end{equation}
The off-diagonal matrix elements describe coherent mixing between
different Pauli components. Equation~\eqref{eq:general_ptm_diagonal}
therefore \tc{recovers the conventional ensemble-averaged Pauli-twirling result~\cite{emerson2007symmetrized,dankert2009exact}, and} gives the standard ensemble-level interpretation of the
protocol: averaging over the random Pauli frames removes the
directional coherence of the error channel and converts it into a
Pauli-diagonal effective process.

\DS{This channel-level result characterizes the ensemble average. Estimating this average experimentally requires sampling multiple
dressed circuits, which may lead to a high compilation and execution cost.
This motivates the practical question of whether a single random realization can already outperform the bare circuit, and under what conditions. To address this question, we next analyze the coherent
error generator of each individual realization and connect it to the target-state fidelity.}

\subsection{Single-realization suppression of coherent \GJ{error} accumulation}
\label{subsec:single_realization}
\label{subsec:covariance_diagnostic}


\tc{Now, we present the complete framework for the single-realization of Pauli-dressed coherent error suppression. }We consider a model \tc{which has weak coherent error} for the \(\ell\)-th selected
two-qubit gate, assuming having the following form:
\begin{equation}
\widetilde U_\ell
=
U_\ell E_\ell,
\qquad
E_\ell
=
e^{-i\epsilon H_\ell},
\qquad
0<\epsilon\ll1,
\label{eq:single_weak_error_model}
\end{equation}
where \(H_\ell\) is a Hermitian two-qubit error generator. Hamiltonian error generators of this form arise naturally in superconducting cross-resonance entangling gates, whose effective dynamics can contain coherent drive, spectator, and unwanted two-qubit Pauli terms
\cite{magesan2020effective,malekakhlagh2020first,sundaresan2020reducing,kaufmann2025characterization,carignandugas2024estimating}. The identity
component of \(H_\ell\) produces only a global phase and is therefore
omitted. We expand the remaining traceless part in the two-qubit
Pauli basis,
\begin{equation}
H_\ell
=
\sum_{\alpha\neq0}
h_{\ell,\alpha}P_\alpha .
\label{eq:single_pauli_expansion}
\end{equation}

For a chosen Pauli dressing \(P_\ell\), the noisy dressed gate is
\begin{align}
\widetilde V_\ell(P_\ell)
&=
Q_\ell\widetilde U_\ell P_\ell
\nonumber\\
&=
Q_\ell U_\ell E_\ell P_\ell
\nonumber\\
&=
U_\ell P_\ell E_\ell P_\ell
\nonumber\\
&=
U_\ell
e^{-i\epsilon P_\ell H_\ell P_\ell}.
\label{eq:single_noisy_dressed_gate}
\end{align}
The ideal logical operation is still exactly \(U_\ell\), whereas the
local coherent generator is transformed as
\begin{equation}
H_\ell
\longrightarrow
H_\ell^{(P_\ell)}
=
P_\ell H_\ell P_\ell .
\label{eq:single_conjugated_generator}
\end{equation}
Since Pauli conjugation preserves each Pauli basis element up to a
sign,
\begin{equation}
P_\ell P_\alpha P_\ell
=
s_{\ell,\alpha}P_\alpha,
\qquad
s_{\ell,\alpha}\in\{+1,-1\},
\label{eq:single_pauli_sign}
\end{equation}
we obtain
\begin{equation}
H_\ell^{(P_\ell)}
=
\sum_{\alpha\neq0}
s_{\ell,\alpha}
h_{\ell,\alpha}P_\alpha .
\label{eq:single_randomized_generator}
\end{equation}
Thus, it randomizes the signs of the Pauli components of the
local generator while leaving the ideal gate unchanged \cite{cai2020mitigating,winick2022concepts}.

Let
\begin{equation}
W_{\ell-1}
=
B_{\ell-1}U_{\ell-1}
\cdots
B_1U_1B_0
\label{eq:single_prefix}
\end{equation}
denote the ideal circuit prefix preceding the \(\ell\)-th selected
entangling gate. \tc{Using Eq.~\eqref{eq:single_weak_error_model}, the local generator can be transformed into such a form as}
\begin{equation}
A_\ell
=
W_{\ell-1}^\dagger
H_\ell
W_{\ell-1}.
\label{eq:single_transported_bare_generator}
\end{equation}
To first order in \(\epsilon\), the bare noisy circuit can be written
as
\begin{equation}
\widetilde U_{\rm bare}
=
U_{\rm ideal}
\exp
\left[
-i\epsilon K_{\rm bare}
+
O\!\left((L\epsilon)^2\right)
\right],
\label{eq:single_bare_effective_unitary}
\end{equation}
where
\begin{equation}
K_{\rm bare}
=
\sum_{\ell=1}^{L}A_\ell .
\label{eq:single_bare_generator}
\end{equation}

For a dressed realization
\(\vec P=(P_1,\ldots,P_L)\), the analogous expression is
\begin{equation}
\widetilde U_{\rm PT}(\vec P)
=
U_{\rm ideal}
\exp
\left[
-i\epsilon K(\vec P)
+
O\!\left((L\epsilon)^2\right)
\right],
\label{eq:single_pt_effective_unitary}
\end{equation}
with
\begin{equation}
K(\vec P)
=
\sum_{\ell=1}^{L}
A_\ell(P_\ell),
\label{eq:single_K_sum}
\end{equation}
and
\begin{equation}
A_\ell(P_\ell)
=
W_{\ell-1}^\dagger
P_\ell H_\ell P_\ell
W_{\ell-1}.
\label{eq:single_layer_generator}
\end{equation}
For a single realization, the sequence \(\vec P\) is sampled once and then held fixed. Using Eq.~\eqref{eq:single_randomized_generator}, this may be written
as
\begin{equation}
A_\ell(P_\ell)
=
\sum_{\alpha\neq0}
s_{\ell,\alpha}h_{\ell,\alpha}
\widetilde P_{\ell,\alpha},
\label{eq:single_transported_expansion}
\end{equation}
where
\begin{equation}
\widetilde P_{\ell,\alpha}
=
W_{\ell-1}^\dagger
P_\alpha
W_{\ell-1}.
\end{equation}
If the prefix \(W_{\ell-1}\) contains non-Clifford layers,
\(\widetilde P_{\ell,\alpha}\) \tc{itself does not need to be} a Pauli operator.
This does not affect the argument, as the sign randomization occurs before the local generator is transported into the common circuit
frame.

Let
\begin{equation}
\ket{\psi_{\rm tar}}
=
U_{\rm ideal}\ket{\psi_0}.
\end{equation}
Expanding the exact circuit overlap to second order in the coherent error strength, the state fidelity can be written as
\begin{align}
F(\vec P)
&=
\left|
\bra{\psi_{\rm tar}}
\widetilde U_{\rm PT}(\vec P)
\ket{\psi_0}
\right|^2
\nonumber\\
&=
\left|
\bra{\psi_0}
e^{-i\epsilon K(\vec P)}
\ket{\psi_0}
\right|^2
+
O\!\left((L\epsilon)^3\right)
\nonumber\\
&=
1
-
\epsilon^2
{\rm Var}_{\psi_0}
\!\left[
K(\vec P)
\right]
+
O\!\left((L\epsilon)^3\right),
\label{eq:single_fidelity_variance}
\end{align}
where
\begin{equation}
{\rm Var}_{\psi_0}(K)
=
\bra{\psi_0}K^2\ket{\psi_0}
-
\bra{\psi_0}K\ket{\psi_0}^2 .
\label{eq:single_variance_definition}
\end{equation}
We assume throughout the perturbative regime in which
\(L\epsilon\ll1\), so that the explicitly displayed second-order term
dominates the omitted higher-order corrections.

Eq.~\eqref{eq:single_fidelity_variance} shows that the state-dependent variance of the total first-order coherent generator is the leading term of the infidelity. To resolve this
variance into layerwise contributions, define
\begin{equation}
\Delta A_\ell(P_\ell)
=
A_\ell(P_\ell)
-
\bra{\psi_0}
A_\ell(P_\ell)
\ket{\psi_0}I
\label{eq:single_centered_generator}
\end{equation}
and the symmetrized connected covariance
\begin{equation}
C_{\ell m}(\vec P)
=
\frac{1}{2}
\bra{\psi_0}
\left\{
\Delta A_\ell(P_\ell),
\Delta A_m(P_m)
\right\}
\ket{\psi_0}.
\label{eq:single_covariance}
\end{equation}
The total variance is then
\begin{align}
{\rm Var}_{\psi_0}
\!\left[
K(\vec P)
\right]
&=
\sum_{\ell=1}^{L}
{\rm Var}_{\psi_0}
\!\left[
A_\ell(P_\ell)
\right]
\nonumber\\
&\quad+
\sum_{\ell\neq m}
C_{\ell m}(\vec P).
\label{eq:single_variance_decomposition}
\end{align}

The first term in Eq.~\eqref{eq:single_variance_decomposition} contains
\(L\) diagonal layerwise contributions. The second term contains the
\(L(L-1)\) cross-layer covariances and therefore determines whether
errors generated at different entangling gates reinforce one another
coherently.

For the bare circuit, \(P_\ell=I\) for all \(\ell\). If the transported
generators act on \(\ket{\psi_0}\) in strongly overlapping directions,
a nonvanishing fraction of the off-diagonal covariances can have
comparable magnitude and the same sign. Under this coherent-alignment
condition,
\begin{equation}
\sum_{\ell\neq m}
C_{\ell m}^{\rm bare}
=
\Theta(L^2),
\label{eq:single_bare_covariance_scaling}
\end{equation}
whereas
\begin{equation}
\sum_{\ell=1}^{L}
{\rm Var}_{\psi_0}(A_\ell)
=
O(L)
\end{equation}
for bounded layerwise error strengths. The cross-layer contribution
then dominates, so that
\begin{equation}
{\rm Var}_{\psi_0}(K_{\rm bare})
=
\Theta(L^2),
\label{eq:single_bare_variance_scaling}
\end{equation}
and the leading bare-circuit infidelity can scale as
\begin{equation}
1-F_{\rm bare}
=
\Theta(L^2\epsilon^2).
\label{eq:single_bare_infidelity}
\end{equation}
The enhanced infidelity is therefore caused by constructive
layer-to-layer covariance.

Now let us turn to the Pauli-dressed realization. To characterize a randomly drawn dressed realization, we take mathematical expectations over the independently sampled Pauli choices at different layers. Because the
identity component of \(H_\ell\) has been removed, uniform averaging over the two-qubit Pauli group gives

\begin{equation}
\mathbb E_{P_\ell}
\left[
P_\ell H_\ell P_\ell
\right]
=
0.
\label{eq:single_local_twirl_zero}
\end{equation}
\tc{by substituting Eq.~\eqref{eq:single_pauli_expansion} into Eq.~\eqref{eq:single_local_twirl_zero}}. Consequently,
\begin{equation}
\mathbb E_{P_\ell}
\left[
A_\ell(P_\ell)
\right]
=
0.
\label{eq:single_zero_mean_layer}
\end{equation}
The Pauli operators are sampled independently at different layers.
Therefore, for \(\ell\neq m\),
\begin{equation}
\mathbb E_{\vec P}
\left[
C_{\ell m}(\vec P)
\right]
=
0.
\label{eq:single_cross_covariance_zero}
\end{equation}
Thus the independently sampled Pauli frames eliminate the expected cross-layer covariance. This result does not require the signs \(s_{\ell,\alpha}\) associated with different Pauli
components within the same layer to be statistically independent.
Only the independence of the Pauli choices at distinct layers and the
zero-mean property in Eq.~\eqref{eq:single_zero_mean_layer} are needed.

Taking the expectation value of Eq.~\eqref{eq:single_variance_decomposition}
therefore gives
\begin{equation}
\mathbb E_{\vec P}
{\rm Var}_{\psi_0}
\!\left[
K(\vec P)
\right]
=
\sum_{\ell=1}^{L}
\mathbb E_{P_\ell}
{\rm Var}_{\psi_0}
\!\left[
A_\ell(P_\ell)
\right].
\label{eq:single_average_variance}
\end{equation}
Since both Pauli conjugation and ideal-frame transport are unitary,
\begin{equation}
{\rm Var}_{\psi_0}
\!\left[
A_\ell(P_\ell)
\right]
\leq
\left\|
A_\ell(P_\ell)
\right\|_\infty^2
=
\|H_\ell\|_\infty^2 .
\label{eq:single_variance_bound}
\end{equation}
If the layerwise coherent error strengths are bounded independently
of the circuit length,
\begin{equation}
\|H_\ell\|_\infty
\leq
h_{\max}
=
O(1),
\label{eq:single_error_bound}
\end{equation}
then
\begin{equation}
\mathbb E_{\vec P}
{\rm Var}_{\psi_0}
\!\left[
K(\vec P)
\right]
\leq
Lh_{\max}^2
=
O(L).
\label{eq:single_pt_variance_scaling}
\end{equation}
Substituting this result into
Eq.~\eqref{eq:single_fidelity_variance} yields
\begin{equation}
\mathbb E_{\vec P}
\left[
1-F(\vec P)
\right]
=
O(L\epsilon^2),
\label{eq:single_pt_infidelity}
\end{equation}
in contrast to the coherent
\(\Theta(L^2\epsilon^2)\) behavior possible in the bare circuit \cite{winick2022concepts,odake2025robust}. 

\DS{We now use this expectation bound to quantify the performance of a single randomly sampled realization.
Since
\({\rm Var}_{\psi_0}[K(\vec P)]\geq0\), Markov's inequality implies
that, for any fixed \(0<\delta<1\),}
\begin{equation}
{\rm Pr}_{\vec P}
\left[
{\rm Var}_{\psi_0}
\!\left[
K(\vec P)
\right]
\geq
\frac{Lh_{\max}^2}{\delta}
\right]
\leq
\delta .
\label{eq:single_markov_bound}
\end{equation}
\DS{Thus, with probability at least \(1-\delta\), a single sampled realization has a generator variance bounded by
\(Lh_{\max}^2/\delta\), which gives an
\(O(L\epsilon^2)\) upper bound on the leading infidelity. The suppression arises from random-sign cancellation among error contributions from different layers within that realization,
analogous to the \(O(\sqrt L)\) typical displacement of a single random walk after \(L\) steps. Averaging the outputs of multiple realizations is therefore not required for this mechanism.
This does not imply that every realization must outperform the bare circuit, particularly in the presence of additional stochastic and single-qubit errors. It does explain, however, why a single realization can already avoid the quadratic enhancement of the infidelity.}

The above analysis also identifies the regime in which two-qubit-gate
PT is expected to be most effective. If the bare transported
generators possess substantial coherently reinforcing off-diagonal
covariance, the dressing removes an \(O(L^2)\) contribution to the
leading infidelity. If the bare covariance is already nearly
diagonal, then
\begin{equation}
{\rm Var}_{\psi_0}(K_{\rm bare})
=
O(L),
\end{equation}
and there is no quadratic coherent contribution for the randomization
to suppress. In Sec.~\ref{sec: covariance}, we evaluate this covariance
structure explicitly for two coherent error models in \tc{a Greenberger–Horne–Zeilinger (GHZ) state} preparation circuit.

Equivalently, the same crossover may be described at the operator
level: coherent alignment can produce a bare first-order generator
whose characteristic size is \(O(L)\), whereas independent Pauli
dressing reduces its typical size in a quadratic operator
norm to \(O(\sqrt L)\). A direct operator-level derivation is given in
Appendix~\ref{app:generator_scaling}.

\tc{
\section{Numerical simulations: covariance diagnostic in the GHZ state preparation}
}

\label{sec:numerics}
\subsection{Scaling of the error variance}
\label{subsec-vark-scaling}

Having established in Sec.~\ref{sec:theory} that the leading contribution to the infidelity is governed by $\operatorname{Var}_{\psi_0}[K(\mathbf P)]$, we first test the resulting depth dependence in a controlled numerical setting. In particular, the perturbative analysis predicts that coherent cross-layer covariance produces $O(L^2)$ growth in the bare circuit, whereas independently sampled PT frames eliminate these cross terms and leave $O(L)$ growth. In this part, we test this accumulation mechanism through the identity-valued mirror circuits on fixed systems of $n=6$ and $n=10$ qubits. These circuits contain only nearest-neighbor CX gates \cite{hashim2021randomized,hashim2023benchmarking}. Defining
\begin{equation}
U_{\mathrm{id}}^{(n)}
=
\mathrm{CX}_{n-2,n-1}\cdots
\mathrm{CX}_{1,2}\mathrm{CX}_{0,1},
\end{equation}
one unit cell is given by
\begin{equation}
B_n
=
\left(U_{\mathrm{id}}^{(n)}\right)^{\dagger}
U_{\mathrm{id}}^{(n)}
=I.
\end{equation}
The unit cell contains $b_n=2(n-1)$ entangling gates, giving $b_6=10$ and $b_{10}=18$. Repeating the cell $r_n$ times produces a circuit with $L=b_n r_n$ CX gates while leaving the ideal evolution equal to the identity. For $n=6$, we use every repetition number from $r_6=1$ to $r_6=54$, giving $L=10,20,\ldots,540$. For $n=10$, we use $r_{10}=1,\ldots,30$, giving $L=18,36,\ldots,540$. 
The input is the product state
\begin{equation}
\ket{\psi_0^{(n)}}=\ket{\phi}^{\otimes n},
\end{equation}
where the Bloch vector of $\ket{\phi}$ is $(1,1,1)/\sqrt{3}$.

\begin{figure*}[t]
    \centering
    \includegraphics[width=0.99\textwidth]{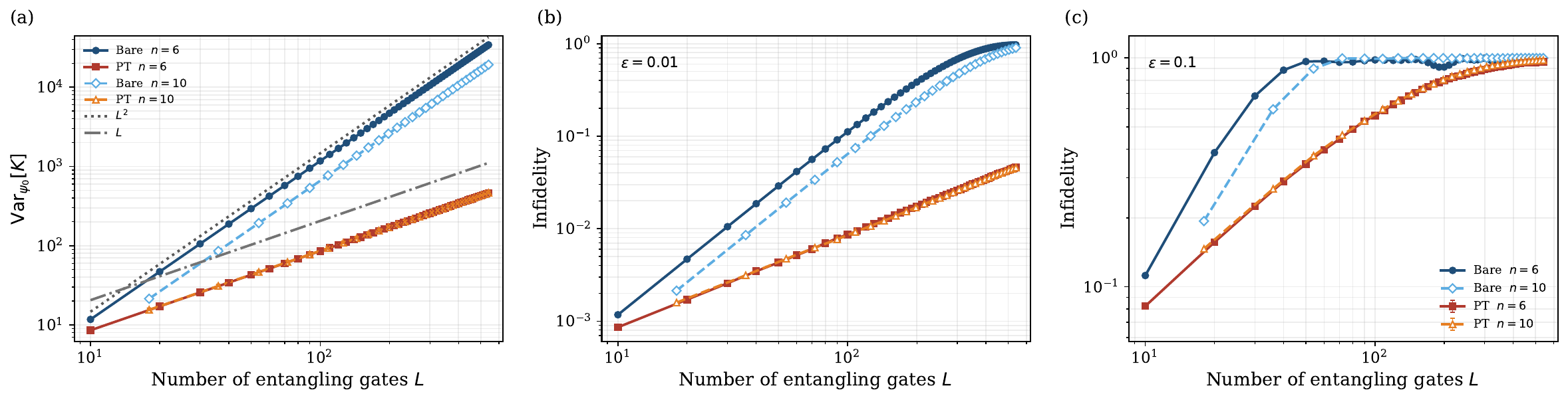}
    \caption{Depth dependence for the six- and ten-qubit mirror circuits. Panel (a) shows the variance of the first-order accumulated error generator. The bare circuits exhibit quadratic growth, whereas the PT results exhibit linear growth. The PT results are obtained from Monte Carlo estimates over 2000 realizations. Panels (b) and (c) show the infidelity $1-F(L,\epsilon)$ at $\epsilon=0.01$ and $\epsilon=0.1$, respectively. Filled symbols with solid lines represent $n=6$, while open symbols with dashed lines represent $n=10$.}
    \label{fig-vark-infidelity-scaling}
\end{figure*}

Following the coherent error model introduced in Sec.~II, the faulty implementation of the $\ell$-th CX gate is
\begin{equation}
\widetilde U_\ell
=
U_\ell e^{-i\epsilon H_{e_\ell}},
\end{equation}
where $e_\ell$ denotes the edge on which the gate acts. Each edge is assigned a coherent two-qubit Hamiltonian containing all 15 nonidentity Pauli components,
\begin{equation}
H_e
=\sum_{Q\in\mathcal P_2\setminus\{II\}}
c_{e,Q}Q,
\qquad
\sum_Q c_{e,Q}^2=1.
\end{equation}
The coefficient vector is sampled from an isotropic Gaussian distribution and normalized. Once sampled, the same $H_e$ is used whenever the corresponding edge appears and throughout all repetitions of the unit cell. The resulting model therefore represents a generic quasi-static coherent error.

With $V_{\ell-1}=U_{\ell-1}\cdots U_1$,
the first-order accumulated generator of the bare circuit is
\begin{equation}
K_{\mathrm{bare}}(L)
=
\sum_{\ell=1}^{L}
V_{\ell-1}^{\dagger}
H_{e_\ell}
V_{\ell-1}.
\end{equation}
Because the ideal unit cell is the identity and the same edge Hamiltonians are reused in every repetition, the transported error pattern is identical in each cell. It follows that
\begin{equation}
K_{\mathrm{bare}}(L)
=
r_n K_{\mathrm{cell}}^{(n)},
\end{equation}
and hence
\begin{equation}
\operatorname{Var}_{\psi_0}
\left[K_{\mathrm{bare}}(L)\right]
=
r_n^2
\operatorname{Var}_{\psi_0}
\left[K_{\mathrm{cell}}^{(n)}\right]
\propto L^2.
\end{equation}

For the PT circuits, an independent two-qubit Pauli $P_\ell$ is sampled uniformly at every CX location, together with the corresponding correction that preserves the ideal gate. The first-order generator for one PT realization $\mathbf P$ is
\begin{equation}
K(\mathbf P)
=
\sum_{\ell=1}^{L}
V_{\ell-1}^{\dagger}
P_\ell H_{e_\ell}P_\ell
V_{\ell-1}.
\end{equation}
Uniform Pauli conjugation changes the signs of the nonidentity Pauli components. The PT eliminates all cross-location contributions after averaging over independently sampled frames, while the diagonal layerwise contributions remain. Therefore
\begin{equation}
\mathbb E_{\mathbf P}
\left[
\operatorname{Var}_{\psi_0}
\left[K(\mathbf P)\right]
\right]
\propto L.
\end{equation}

Fig.~\ref{fig-vark-infidelity-scaling}(a) shows the resulting scaling. Fits to $\operatorname{Var}(K)=aL^p$ give
\[
p_{\mathrm{bare}}^{(6)}
=
p_{\mathrm{bare}}^{(10)}
=2.000,
\]
for the exact bare results, and
\[
p_{\mathrm{bare}}^{(6)}
=1.003,
\qquad
p_{\mathrm{bare}}^{(10)}
=1.002,
\]
for the sampled PT realizations, confirming that after applying the two-qubit Pauli twirling, the quadratic growth in the bare circuit is suppressed to linear growth.

We then examine the exact output-state infidelity $\mathcal I(L,\epsilon)
= 1-F(L, \epsilon)$ at different perturbation strengths.
For fixed $L$ and sufficiently small $\epsilon$, perturbation theory gives
\begin{equation}
\mathcal I(L,\epsilon)
=
\epsilon^2
\operatorname{Var}_{\psi_0}[K(L)]
+O(\epsilon^3).
\end{equation}
Panels (b) and (c) of Fig.~\ref{fig-vark-infidelity-scaling} show the results for $\epsilon=0.01$ and $\epsilon=0.1$, respectively. For both amplitudes and both system sizes, the PT-averaged infidelity remains below the corresponding bare result at every sampled depth. The PT-averaged results remain close to linear growth over the displayed range, while the bare circuit infidelity shows a nearly quadratic growth and reaches $1$ at larger $L$. Comparing the two amplitudes therefore illustrates how suppression of the first-order error variance delays fidelity loss, which is consistent with the theory in Sec.~\ref{sec:theory}.

\subsection{Connected covariance as a diagnostic of PT performance}
\label{sec: covariance}
Having verified the predicted depth scaling of the first-order error variance, we now examine the general criteria that determine when a given circuit can benefit from PT. The connected covariance matrix introduced in Sec.~\ref{subsec:single_realization} resolves the leading infidelity into individual layerwise contributions and interference between errors generated at different entangling gates. A substantial constructive off-diagonal contribution identifies coherent accumulation that PT can suppress, whereas a nearly diagonal matrix indicates little cross-layer accumulation to remove. We use GHZ-state preparation \cite{greenberger1989going} to demonstrate these two covariance regimes in a concrete circuit.

We consider a fan-like GHZ-preparation circuit acting on an even number
\(N\) of qubits. Starting from \(\ket{0}^{\otimes N}\), a Hadamard gate
\(H_c\) is first applied to the left qubit of the central pair, and the
resulting coherence is spread outward by a sequence of CNOT gates along
a growing tree \cite{mooney2021generation}. The ideal circuit is
\begin{equation}
U_{\rm GHZ}
=
U_LU_{L-1}\cdots U_2U_1H_c,
\label{eq:ghz_circuit}
\end{equation}
where each \(U_\ell\) is a CNOT gate and \(L=N-1\) for the circuit
geometry considered here. The target state is
\begin{equation}
\ket{\psi_{\rm tar}}
=
U_{\rm GHZ}\ket{0}^{\otimes N}
=
\ket{\mathrm{GHZ}_N}
=
\frac{
\ket{0}^{\otimes N}
+
\ket{1}^{\otimes N}
}{\sqrt{2}}.
\label{eq:ghz_target}
\end{equation}

The GHZ-state fidelity can be expressed in terms of the stabilizer group
\begin{equation}
\mathcal S_{\rm GHZ}
=
\left\langle
X^{\otimes N},
Z_1Z_2,
Z_2Z_3,
\ldots,
Z_{N-1}Z_N
\right\rangle .
\label{eq:ghz_stabilizer_group}
\end{equation}
The projector onto the target state can be written as \cite{gottesman1997stabilizer}
\begin{equation}
\Pi_{\rm GHZ}
=
\ket{\mathrm{GHZ}_N}
\!\bra{\mathrm{GHZ}_N}
=
\frac{1}{2^N}
\sum_{S\in\mathcal S_{\rm GHZ}}S,
\label{eq:ghz_projector}
\end{equation}
thus the fidelity of a prepared state \(\rho\) is
\begin{equation}
F_{\rm GHZ}
=
\bra{\mathrm{GHZ}_N}
\rho
\ket{\mathrm{GHZ}_N}
=
\frac{1}{2^N}
\sum_{S\in\mathcal S_{\rm GHZ}}
{\rm Tr}(\rho S).
\label{eq:ghz_stabilizer_fidelity}
\end{equation}
Equation~\eqref{eq:ghz_stabilizer_fidelity} motivates the
stabilizer-sampling estimator used in the hardware experiments. In the numerical simulations below, the fidelity is evaluated directly from
the simulated final state \cite{flammia2011direct,dasilva2011practical,pallister2018optimal}. Here, the numerical simulations include fake-backend background noise and readout errors in addition to the imposed coherent errors.

To model coherent two-qubit control errors, the ideal CNOT gate in each
entangling layer is replaced by
\begin{equation}
\widetilde U_\ell
=
U_\ell
e^{-i\epsilon H_\ell},
\label{eq:numerical_error_model}
\end{equation}
where \(\epsilon\) controls the coherent error strength. As illustrative test cases, we use two local two-qubit Pauli generators,
\begin{equation}
H_\ell^{(ZZ)}
=
Z_{c_\ell}Z_{t_\ell},
\qquad
H_\ell^{(XX)}
=
X_{c_\ell}X_{t_\ell},
\label{eq:zz_xx_generators}
\end{equation}
where \(c_\ell\) and \(t_\ell\) denote the control and target qubits of
the \(\ell\)-th CNOT gate.
These generators are chosen to illustrate different covariance structures within this GHZ-preparation circuit.

\begin{figure*}[tp]
\centering
\includegraphics[width=0.95\textwidth]{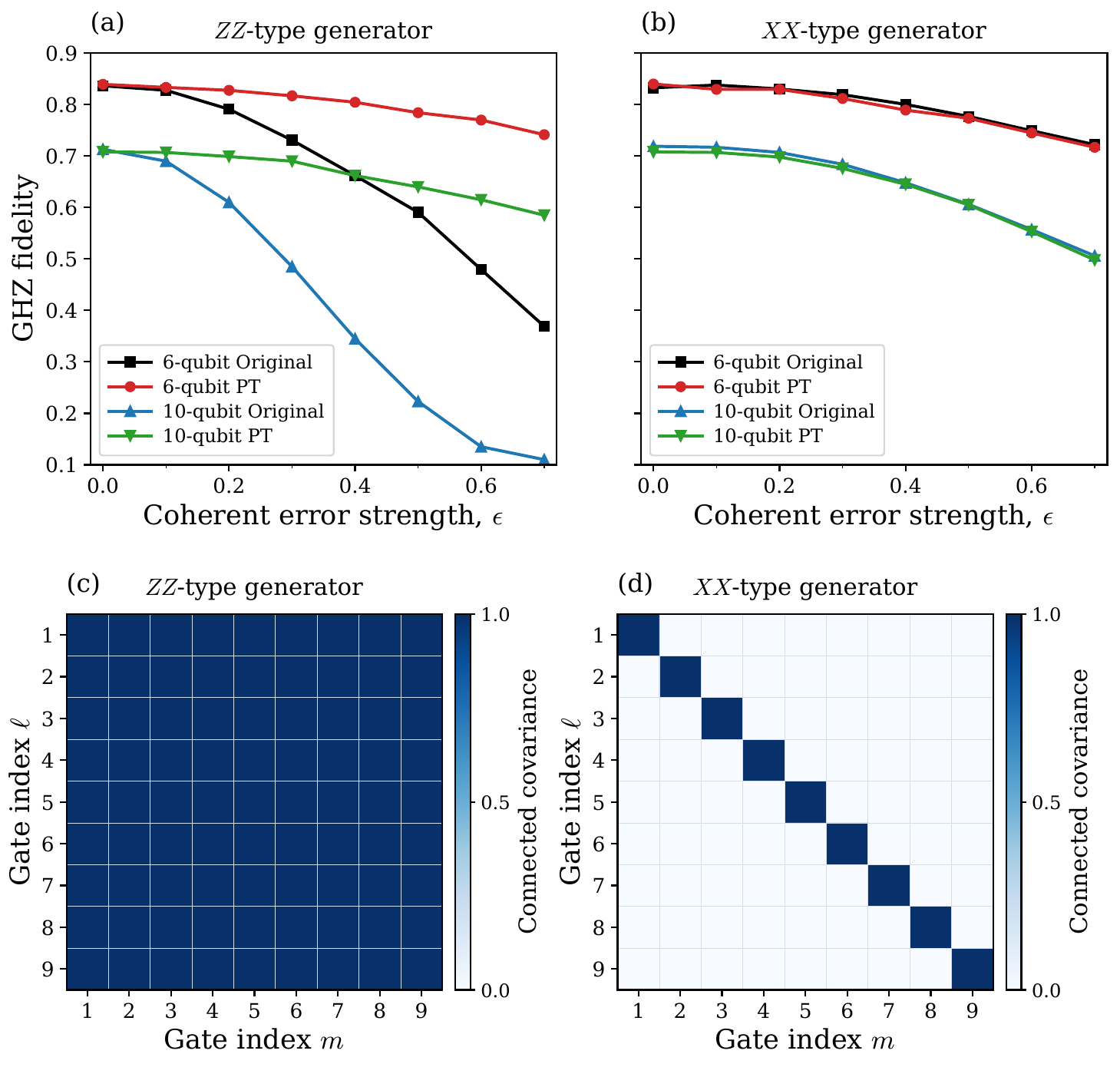}
\caption{
Demonstration of the connected covariance diagnostic in the fan-like GHZ-preparation circuit.
Panels (a) and (b) show the
fidelity as a function of the coherent error strength \(\epsilon\) for
the \(ZZ\)-type and \(XX\)-type generators, respectively. In each panel,
the bare GHZ-preparation circuit is compared with its PT counterpart
for \(N=6\) and \(N=10\). In (a) and (b), the simulations include fake-backend background noise and readout errors. Panels (c) and (d) show the corresponding connected covariance matrices $C_{\ell m}$ for $N=10$. In this circuit, the constructive off-diagonal covariance in panel (c) predicts the pronounced PT improvement in panel (a), while the diagonal matrix in panel (d) predicts the small change in panel (b). Note that the $ZZ$ and $XX$ generators illustrate these two regimes only for the specified circuit and input state, and their contrasting responses to PT do not imply a universal distinction between the two error types.
}
\label{fig:numerical_ghz}
\end{figure*}

To evaluate the diagnostic for these examples, we examine the transported first-order generators. Let
\begin{equation}
W_{\ell-1}
=
U_{\ell-1}\cdots U_2U_1H_c,
\qquad
W_0=H_c,
\label{eq:ghz_prefix}
\end{equation}
be the ideal circuit prefix preceding the \(\ell\)-th CNOT. In the
input frame, the transported generator is
\begin{equation}
A_\ell
=
W_{\ell-1}^\dagger
H_\ell
W_{\ell-1}.
\label{eq:ghz_transported_generator_input}
\end{equation}
The bare first-order generator is consequently
\begin{equation}
K_{\rm bare}
=
\sum_{\ell=1}^{L}A_\ell .
\label{eq:ghz_bare_generator}
\end{equation}

For visualization, it is convenient to use the equivalent output-frame
generators
\begin{equation}
G_\ell
=
U_{\rm GHZ}
A_\ell
U_{\rm GHZ}^\dagger .
\label{eq:ghz_transported_generator_output}
\end{equation}
Since
\(
\ket{\mathrm{GHZ}_N}
=
U_{\rm GHZ}\ket{0}^{\otimes N}
\),
all moments are unchanged by this change of frame. For example,
\begin{equation}
\bra{0}^{\otimes N}
A_\ell A_m
\ket{0}^{\otimes N}
=
\bra{\mathrm{GHZ}_N}
G_\ell G_m
\ket{\mathrm{GHZ}_N}.
\label{eq:ghz_frame_equivalence}
\end{equation}
We therefore evaluate the covariance matrix in the output frame, where
its action can be interpreted directly on the target GHZ state.

Defining
\begin{equation}
\langle O\rangle_{\rm GHZ}
=
\bra{\mathrm{GHZ}_N}
O
\ket{\mathrm{GHZ}_N},
\end{equation}
the symmetrized overlap matrix is
\begin{equation}
M_{\ell m}
=
\frac{1}{2}
\left\langle
G_\ell G_m+G_mG_\ell
\right\rangle_{\rm GHZ},
\label{eq:ghz_overlap_matrix}
\end{equation}
and the covariance matrix is
\begin{equation}
C_{\ell m}
=
M_{\ell m}
-
\langle G_\ell\rangle_{\rm GHZ}
\langle G_m\rangle_{\rm GHZ}.
\label{eq:ghz_covariance_matrix}
\end{equation}
Equivalently, if
\begin{equation}
\Delta G_\ell
=
G_\ell
-
\langle G_\ell\rangle_{\rm GHZ}I,
\end{equation}
then
\begin{equation}
C_{\ell m}
=
\frac{1}{2}
\left\langle
\left\{
\Delta G_\ell,\Delta G_m
\right\}
\right\rangle_{\rm GHZ}.
\label{eq:ghz_covariance_centered}
\end{equation}
This definition makes \(C_{\ell m}\) real and symmetric and is the
output-frame version of the covariance introduced in
Sec.~\ref{subsec:single_realization}.

The leading variance of the total bare generator is
\begin{equation}
{\rm Var}_{\rm GHZ}
\left(
\sum_{\ell=1}^{L}G_\ell
\right)
=
\sum_{\ell,m=1}^{L}
C_{\ell m}.
\label{eq:ghz_variance_covariance}
\end{equation}
The diagonal terms \(C_{\ell\ell}\) quantify the individual
contribution of each entangling layer. The off-diagonal terms
\(C_{\ell m}\), with \(\ell\neq m\), quantify the signed overlap between the error directions generated at different layers.

This interpretation can also be understood by defining
the layerwise leakage vectors
\begin{equation}
\ket{\chi_\ell}
=
\Delta G_\ell
\ket{\mathrm{GHZ}_N}.
\label{eq:ghz_leakage_vector}
\end{equation}
The covariance matrix then satisfies
\begin{equation}
C_{\ell m}
=
{\rm Re}
\braket{\chi_\ell|\chi_m}.
\label{eq:ghz_covariance_geometric}
\end{equation}
Positive off-diagonal entries describe constructive interference between layerwise error directions, while negative entries describe destructive interference. Entries close to zero indicate little pairwise contribution to the leading infidelity. The sign and magnitude of the total off-diagonal contribution therefore determine the coherent accumulation available for PT to suppress.

For the Pauli error models in Eq.~\eqref{eq:zz_xx_generators}, Pauli conjugation changes only the signs of the transported generators. Independent dressing leaves the diagonal variances unchanged and removes the off-diagonal contribution on average, as shown in Sec.~\ref{subsec:single_realization}. A substantial net positive sum of off-diagonal entries consequently identifies a favorable regime for PT. When this sum grows as $L^2$ while the diagonal sum grows as $L$, Eq.~\eqref{eq:single_markov_bound} predicts suppression of the quadratic enhancement in a randomly sampled realization at any fixed confidence level. If the covariance matrix is nearly diagonal, there is little cross-layer coherent accumulation to suppress, and correspondingly little improvement is expected in the leading infidelity.

The covariance matrices in Fig.~\ref{fig:numerical_ghz}(c) and (d) illustrate these two regimes. In the specified fan-like GHZ circuit, the $ZZ$-type generator produces substantial constructive off-diagonal covariance, so errors generated at different CNOT layers reinforce one another. The $XX$-type generator instead produces a diagonal matrix, with little coherent reinforcement between layers. The diagnostic therefore predicts a pronounced PT advantage for the former covariance structure and little improvement for the latter. The fidelity results in Fig.~\ref{fig:numerical_ghz}(a) and (b) are consistent with these predictions for both $N=6$ and $N=10$. For the case with constructive off-diagonal covariance, the bare-circuit fidelity decreases rapidly as $\epsilon$ increases, whereas the PT-dressed circuit retains a substantially higher fidelity over the full range considered. For the nearly diagonal case, the bare and PT-dressed fidelity curves remain close.

These simulations demonstrate the general covariance criterion in a concrete state-preparation task, showing that we can predict the PT protocol's effectiveness through the connected covariance matrices for the specified circuit and input state. Changing the circuit or the input can change this covariance matrix for the same local error generator, therefore the $ZZ$ and $XX$ examples used here only illustrate the two regimes of the criterion in this particular case, and their contrasting responses to PT do not imply a universal distinction between the two error types.

\section{Hardware benchmarks on IBM quantum processors}
\label{sec:hardware}

We now investigate the performance of the two-qubit-gate PT on IBM superconducting quantum processors \cite{ware2021experimental,hashim2021randomized,hashim2023benchmarking}. The purpose of this section is to test whether the coherent-error suppression mechanism established
in Sec.~\ref{sec:theory} remains effective under realistic hardware
noise. We consider two circuit families with different
structures and task-level observables. First, we study the fan-like
GHZ-state preparation circuit used in the numerical simulations.
Second, we consider a Trotterized transverse-field Ising circuit, which tests the protocol in a repeated time-evolution setting. An additional quantum Fourier transform (QFT) benchmark, which probes the input dependence of the PT performance, is presented in Appendix~\ref{app:hardware_qft}.

For each circuit instance, we compare the bare implementation with
\(M\) independently sampled PT realizations \cite{fruitwala2024hardware},
\begin{equation}
C_{\rm PT}^{(1)},
C_{\rm PT}^{(2)},
\ldots,
C_{\rm PT}^{(M)} .
\label{eq:hardware_pt_ensemble}
\end{equation}
All realizations implement exactly the same ideal circuit since every
Pauli dressing is accompanied by the matched correction introduced in
Sec.~\ref{sec:theory}. 

\tc{
\subsection{The GHZ state preparation}
}
\label{subsec:hardware_ghz}

We first implement the fan-like GHZ-preparation circuits introduced in
Sec.~\ref{sec:numerics} \cite{song2017ten,wei2020verifying,mooney2021generation}. Similar as above, for each system size \(N\), using the GHZ stabilizer group $\mathcal S_{\rm GHZ} = \left\langle X^{\otimes N}, Z_1Z_2, Z_2Z_3, \ldots, Z_{N-1}Z_N \right\rangle$, the target-state fidelity is
\begin{equation}
F_{\rm GHZ}
=
\frac{1}{2^N}
\sum_{S\in\mathcal S_{\rm GHZ}}
{\rm Tr}(\rho S).
\label{eq:hardware_ghz_fidelity}
\end{equation}
In the experiment, stabilizer elements are sampled uniformly from
\(\mathcal S_{\rm GHZ}\), and their Pauli expectation values are
measured on the prepared state. If
\(S_1,\ldots,S_{N_s}\) are the sampled stabilizers, the corresponding
Monte Carlo estimator is \cite{flammia2011direct,dasilva2011practical}
\begin{equation}
\widehat F_{\rm GHZ}
=
\frac{1}{N_s}
\sum_{a=1}^{N_s}
\widehat{\langle S_a\rangle}.
\label{eq:hardware_ghz_sampling_estimator}
\end{equation}
Uniform stabilizer sampling makes
Eq.~\eqref{eq:hardware_ghz_sampling_estimator} an unbiased estimator of
Eq.~\eqref{eq:hardware_ghz_fidelity}, apart from hardware and
measurement errors.

\begin{figure}[t]
\centering
\includegraphics[width=\columnwidth]{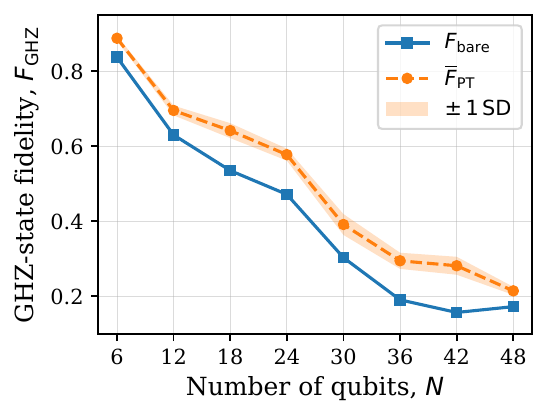}
\caption{
Measured GHZ-state fidelity as a function of the number of qubits
\(N\) on an IBM superconducting quantum processor. The bare fan-like
GHZ-preparation circuit is compared with the mean fidelity over
\(M\) independently sampled PT realizations. The shaded band shows
the mean PT fidelity plus or minus one standard deviation. The PT fidelity
and the lower edge of this band remain above the bare-circuit
fidelity throughout the tested range.
}
\label{fig:hardware_ghz}
\end{figure}

For the bare circuit, the measured fidelity is denoted by
\(F_{\rm bare}\). For the \(m\)-th PT realization, we denote the
estimated fidelity by \(\widehat F_{\rm PT}^{(m)}\). The mean PT
fidelity is
\begin{equation}
\overline F_{\rm PT}
=
\frac{1}{M}
\sum_{m=1}^{M}
\widehat F_{\rm PT}^{(m)}.
\label{eq:hardware_ghz_pt_average}
\end{equation}
In Fig.~\ref{fig:hardware_ghz}, the shaded band represents one
standard deviation above and below this mean.

Both fidelities generally decrease with increasing \(N\), as larger
GHZ-preparation circuits require more two-qubit gates. Nevertheless,
the PT fidelity remains above the bare-circuit fidelity throughout
the tested range. For example, at \(N=24\), the bare circuit gives
\(F_{\rm bare}=0.472\), whereas the PT-averaged fidelity reaches
\(\overline F_{\rm PT}=0.578\). At \(N=36\), the corresponding values
are \(F_{\rm bare}=0.190\) and \(\overline F_{\rm PT}=0.295\).
The lower edge of the one-standard-deviation band also remains above the
bare-circuit fidelity at every measured system size.

\DS{The systematic improvement of
\(\overline F_{\rm PT}\) indicates that the device noise contains a coherent component whose accumulation can be reduced by randomizing the Pauli frame of the entangling gates. The standard-deviation band
remains above the bare-circuit fidelity across all tested system
sizes, indicating that the PT improvement is robust to variations
among individual realizations. This is consistent with the single-realization mechanism of Sec.~\ref{subsec:single_realization},
although the hardware data alone do not identify the complete
microscopic noise generator.}

\subsection{Trotterized Hamiltonian simulation}
\label{subsec:hardware_tfim}

\begin{figure*}[t]
\centering
\includegraphics[width=\textwidth]{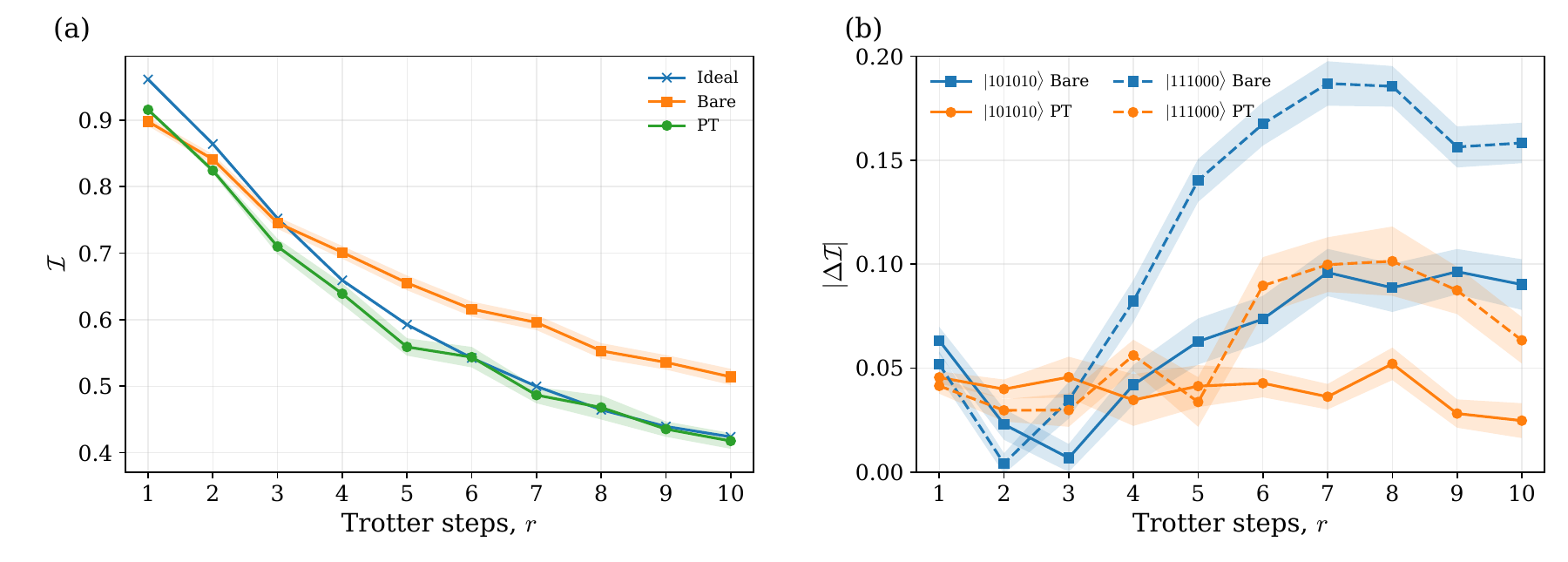}
\caption{
Imbalance dynamics in the Trotterized transverse-field Ising benchmark
on an IBM Pittsburgh processor, with \(N=6\), \(J=1\), \(h=0.7\),
and \(\Delta t=0.2\).
(a) Imbalance \(\mathcal I(r)\) for the initial state
\(\ket{101010}\), compared with the noiseless Trotter circuit.
(b) Imbalance error \(|\Delta\mathcal I(r)|\) for
\(\ket{101010}\) and \(\ket{111000}\), shown by solid and dashed curves, respectively.
The PT curve shows the mean value over many sampled realizations, \DS{where the absolute error is evaluated for each realization before averaging}. Bare denotes the circuit without randomization.
Shaded bands indicate $1\sigma$ standard deviation after bootstrapping the data.
Complementary profile-distance diagnostics
are presented in Appendix~\ref{app:hardware_tfim_profiles}.
}
\label{fig:tfim_hardware}
\end{figure*}

Next, we test the protocol in a digital Hamiltonian-simulation circuit \cite{smith2019simulating,kim2023evidence,lloyd1996universal}. We consider the one-dimensional transverse-field Ising Hamiltonian
\begin{equation}
H
=
J\sum_{j=0}^{N-2}Z_jZ_{j+1}
+
h\sum_{j=0}^{N-1}X_j .
\label{eq:tfim_hamiltonian_hardware}
\end{equation}
Here, $N$ denotes the chain length, $J$ is the coupling strength, and $h$ stands for the transverse-field strength. For a total evolution time \(t=r\Delta t\), we use the first-order Trotterization \cite{trotter1959product,suzuki1976generalized,childs2021theory}
\begin{equation}
U_{\rm Trot}(t)
=
\left[
\prod_{j=0}^{N-1}
e^{-i\Delta t h X_j}
\prod_{j=0}^{N-2}
e^{-i\Delta t J Z_jZ_{j+1}}
\right]^r ,
\label{eq:tfim_trotter_hardware}
\end{equation}
where \(r\) denotes the number of Trotter steps. We consider two computational-basis initial states, $|101010\rangle$ and $|111000\rangle$. The first is a staggered pattern, and the second is a domain-wall-like configuration. 

Each nearest-neighbor interaction block is decomposed as
\begin{equation}
e^{-i\theta Z_jZ_{j+1}}
=
CX_{j,j+1}
R_z(2\theta)_{j+1}
CX_{j,j+1},
\qquad
\theta=J\Delta t .
\label{eq:rzz_decomposition_hardware}
\end{equation}
PT is applied only to the CNOT gates in this decomposition. Thus every PT-dressed circuit remains logically equivalent to the same ideal Trotterized time-evolution circuit in the absence of noise.

\tc{To illustrate the effect of PT on the time evolution of the many-body Hamiltonian,} we use the imbalance relative to the
initial spin pattern,
\begin{equation}
\mathcal I(t)
=
\frac{1}{N}
\sum_{j=0}^{N-1}
z_j(0)\langle Z_j(t)\rangle,
\label{eq:imbalance_hardware}
\end{equation}
where \(z_j(0)=+1\) for an initial \(\ket{0}\) and
\(z_j(0)=-1\) for an initial \(\ket{1}\). The imbalance measures
the memory of the initial spatial spin pattern under time evolution, \tc{which acts as a direct, experimentally measurable proxy for quantum memory retention and ergodicity~\cite{PhysRevB.95.014202}. }
We quantify its deviation from the ideal Trotterized dynamics by
\begin{equation}
|\Delta\mathcal I(r)|
=
\left|
\mathcal I_{\rm exp}(r)-\mathcal I_{\rm id}(r)
\right|,
\label{eq:imbalance_error_hardware}
\end{equation}
where \(\mathcal I_{\rm exp}(r)\) is the experimental imbalance and
\(\mathcal I_{\rm id}(r)\) is obtained from the noiseless Trotter
circuit. These quantities are obtained from computational-basis
measurements and the corresponding ideal simulation.

Figure~\ref{fig:tfim_hardware} compares the bare circuit with PT.
Fig.~\ref{fig:tfim_hardware}(a) shows the imbalance dynamics for
\(\ket{101010}\). The bare circuit increasingly deviates from the
ideal curve at larger \(r\), while the mean PT imbalance remains
much closer to the ideal result at larger depths.
Fig.~\ref{fig:tfim_hardware}(b) compares the bare-circuit absolute
error with the PT mean absolute error for both initial states.
The improvement is especially evident
at larger \(r\). At \(r=10\), the error for \(\ket{101010}\)
decreases from approximately \(9.02\times10^{-2}\) for the bare
circuit to \(2.47\times10^{-2}\) for PT. For \(\ket{111000}\),
the corresponding error decreases from approximately
\(1.58\times10^{-1}\) to \(6.34\times10^{-2}\).
The shaded bands represent the statistical uncertainty arising from the finite-shot measurements on the quantum hardware, which affects both the bare and PT results. \tc{In addition to the above, to test whether this improvement extends beyond the imbalance, we have also examined the full local
magnetization profile and the nearest-neighbor correlation profile
for the same circuits and initial states (see Appendix~\ref{app:hardware_tfim_profiles}). These complementary
diagnostics also show reduced errors at larger Trotter depths.}

Together, these results extend the GHZ benchmark to time-evolution
circuits built from repeated local Hamiltonian-simulation blocks.
The reduction of observable errors at larger Trotter depths is
consistent with the suppression of a coherent error component accumulated across the repeated \(ZZ\)-interaction blocks, showing that our PT randomization can effectively suppress noise across different circuit structures.

\section{Conclusion}
\label{sec:conclusion}

In this work, we have introduced a single-realization framework for standard PT in circuits containing many two-qubit Clifford entangling gates. Each entangling gate is dressed by a random two-qubit Pauli and a matched Pauli correction, so that every random realization implements exactly the same ideal circuit.  In each single-realization circuit, the method randomizes the signs of coherent Pauli error components and can convert aligned coherent accumulation into random-walk-like accumulation.

A central contribution of this work is the covariance-based diagnostic for when PT is expected to be effective. Numerical simulations of a fan-like GHZ circuit show that a \(ZZ\)-type coherent error produces large off-diagonal covariance between transported error generators and is strongly suppressed by PT, whereas an \(XX\)-type coherent error produces an almost diagonal covariance matrix and shows little PT advantage. This identifies the coherent-noise structures that can be effectively suppressed by our PT methods.

Hardware experiments further show that the two-qubit-gate PT improves a broad set of structured circuits. It enhances multiqubit GHZ-state fidelity over a wide range of system sizes and reduces observable errors in Trotterized transverse-field Ising dynamics. The additional QFT benchmark in Appendix~\ref{app:hardware_qft} illustrates the input dependence of the PT improvement. These results support the view that this PT method is a practical coherent-error suppression method for structured quantum circuits whose ideal dynamics rely on repeated nontrivial two-qubit operations.

Several directions merit further investigation. One is to combine PT with quantum error correction and examine how suppressing coherent accumulation in syndrome-extraction circuits affects logical error rates, as well as whether syndrome information can guide subsequent Pauli choices. Another is to use knowledge of the dominant noise generators and their propagation through the circuit to design restricted Pauli sets or nonuniform sampling distributions, rather than sampling all Pauli dressings with equal probability. Such strategies would require identifying which sampling conditions preserve the cancellation of the relevant coherent error contributions. The covariance criterion developed here could also guide selective dressing of the entangling layers that contribute most strongly to coherent enhancement, providing a framework for exploring the trade-off between coherent error suppression and the number of additional Pauli operations.

\appendix

\section{Scaling of the effective coherent generator}
\label{app:generator_scaling}

The main text derives the fidelity improvement directly from the scaling of
\(\operatorname{Var}_{\psi_0}[K(\vec P)]\), which is the leading term of infidelity. In this appendix, we present a complementary operator-level description. We show that coherent alignment can produce an \(O(L)\) bare first-order generator, whereas independent Pauli dressing reduces the root-mean-square size of the dressed generator to \(O(\sqrt L)\) in a quadratic operator norm.

We use the layer generators introduced in Sec.~\ref{subsec:single_realization},
\begin{equation}
A_\ell
=
W_{\ell-1}^\dagger H_\ell W_{\ell-1},
\qquad
K_{\rm bare}
=
\sum_{\ell=1}^{L}A_\ell,
\label{eq:app_bare_K}
\end{equation}
and
\begin{equation}
A_\ell(P_\ell)
=
W_{\ell-1}^\dagger
P_\ell H_\ell P_\ell
W_{\ell-1},
\qquad
K(\vec P)
=
\sum_{\ell=1}^{L}A_\ell(P_\ell).
\label{eq:app_pt_K}
\end{equation}

For definiteness, we quantify the operator size using the normalized Hilbert--Schmidt inner product
\begin{equation}
\langle A,B\rangle_2
=
\frac{1}{d}
\operatorname{Tr}(A^\dagger B),
\qquad
d=2^n,
\label{eq:app_hs_inner_product}
\end{equation}
and the associated norm
\begin{equation}
\|A\|_2
=
\sqrt{\langle A,A\rangle_2}.
\label{eq:app_hs_norm}
\end{equation}
The normalization removes the trivial Hilbert-space-dimension factor.

\subsection{Coherent growth in the bare circuit}

Let \(\{M_\beta\}\) be a complete orthonormal Hermitian operator basis with respect to Eq.~\eqref{eq:app_hs_inner_product}. Expand each transported generator as
\begin{equation}
A_\ell
=
\sum_\beta c_{\ell,\beta}M_\beta,
\label{eq:app_bare_basis_expansion}
\end{equation}
where the coefficients \(c_{\ell,\beta}\) are real. The total bare generator is
\begin{equation}
K_{\rm bare}
=
\sum_\beta
\left(
\sum_{\ell=1}^{L}c_{\ell,\beta}
\right)
M_\beta,
\label{eq:app_bare_basis_sum}
\end{equation}
and hence
\begin{equation}
\|K_{\rm bare}\|_2^2
=
\sum_\beta
\left|
\sum_{\ell=1}^{L}c_{\ell,\beta}
\right|^2 .
\label{eq:app_bare_norm}
\end{equation}

Suppose that there exists a dominant operator direction \(M_{\beta_*}\) for which the coefficients remain aligned over a nonvanishing fraction of the circuit. More precisely, suppose that
\begin{equation}
c_{\ell,\beta_*}=a_\ell+\delta_\ell,
\label{eq:app_aligned_coefficients}
\end{equation}
where the \(a_\ell\) have a common sign and \(O(1)\) magnitude for \(O(L)\) layers, while
\begin{equation}
\sum_{\ell=1}^{L}\delta_\ell=o(L).
\end{equation}
Then
\begin{equation}
\left|
\sum_{\ell=1}^{L}c_{\ell,\beta_*}
\right|
=
\Theta(L),
\label{eq:app_aligned_sum}
\end{equation}
where $\Theta(L)$ denotes asymptotically linear scaling with $L$. Equation~\eqref{eq:app_bare_norm} therefore implies the lower bound
\begin{equation}
\|K_{\rm bare}\|_2
\geq
\left|
\sum_{\ell=1}^{L}c_{\ell,\beta_*}
\right|
=
\Theta(L).
\label{eq:app_bare_lower_bound}
\end{equation}
On the other hand, if the individual layer generators have bounded norm,
\begin{equation}
\|A_\ell\|_2=O(1),
\end{equation}
the triangle inequality gives
\begin{equation}
\|K_{\rm bare}\|_2
\leq
\sum_{\ell=1}^{L}\|A_\ell\|_2
=
O(L).
\label{eq:app_bare_upper_bound}
\end{equation}
Combining Eqs.~\eqref{eq:app_bare_lower_bound} and
\eqref{eq:app_bare_upper_bound}, we obtain
\begin{equation}
\|K_{\rm bare}\|_2
=
\Theta(L)
\label{eq:app_bare_linear_scaling}
\end{equation}
whenever the circuit transports a repeated coherent error into some common dominant operator directions.

The same alignment condition can also be expressed through the action of the generators on the input state. If
\begin{equation}
A_\ell\ket{\psi_0}
=
a_\ell\ket{\phi}
+
\ket{r_\ell},
\label{eq:app_state_alignment}
\end{equation}
where \(\ket{\phi}\) is a common error direction, the \(a_\ell\) have approximately fixed sign, and the accumulated remainder satisfies
\begin{equation}
\left\|
\sum_{\ell=1}^{L}\ket{r_\ell}
\right\|
=
o(L),
\end{equation}
then
\begin{equation}
K_{\rm bare}\ket{\psi_0}
=
\left(
\sum_{\ell=1}^{L}a_\ell
\right)
\ket{\phi}
+
\sum_{\ell=1}^{L}\ket{r_\ell},
\label{eq:app_state_alignment_sum}
\end{equation}
and therefore
\begin{equation}
\|K_{\rm bare}\ket{\psi_0}\|
=
\Theta(L).
\label{eq:app_state_linear_scaling}
\end{equation}
This state-level statement is the geometric meaning of coherent accumulation: errors generated at different layers drive the state in approximately the same direction.

\subsection{Root-mean-square scaling under Pauli dressing}

For the dressed circuit, the local generator is
\begin{equation}
A_\ell(P_\ell)
=
W_{\ell-1}^\dagger
P_\ell H_\ell P_\ell
W_{\ell-1}.
\end{equation}
Using the Pauli expansion of \(H_\ell\), it can be written as
\begin{equation}
A_\ell(P_\ell)
=
\sum_{\alpha\neq0}
s_{\ell,\alpha}
h_{\ell,\alpha}
\widetilde P_{\ell,\alpha},
\label{eq:app_pt_sign_expansion}
\end{equation}
where
\begin{equation}
\widetilde P_{\ell,\alpha}
=
W_{\ell-1}^\dagger P_\alpha W_{\ell-1}.
\end{equation}
The signs associated with different Pauli components within one layer need not be independent. The derivation below requires only that the Pauli operators \(P_\ell\) are sampled independently at different layers.

Since the identity component of \(H_\ell\) has been removed, uniform Pauli conjugation gives
\begin{equation}
\mathbb E_{P_\ell}
\left[
A_\ell(P_\ell)
\right]
=
0.
\label{eq:app_zero_mean}
\end{equation}
We now evaluate the mean squared norm of the total dressed generator:
\begin{align}
\mathbb E_{\vec P}
\|K(\vec P)\|_2^2
&=
\mathbb E_{\vec P}
\left\langle
\sum_{\ell=1}^{L}A_\ell(P_\ell),
\sum_{m=1}^{L}A_m(P_m)
\right\rangle_2
\nonumber\\
&=
\sum_{\ell=1}^{L}
\mathbb E_{P_\ell}
\|A_\ell(P_\ell)\|_2^2
\nonumber\\
&\quad+
\sum_{\ell\neq m}
\mathbb E_{\vec P}
\left[
\langle
A_\ell(P_\ell),
A_m(P_m)
\rangle_2
\right].
\label{eq:app_pt_norm_expansion}
\end{align}
For \(\ell\neq m\), independence of \(P_\ell\) and \(P_m\), together with Eq.~\eqref{eq:app_zero_mean}, gives
\begin{align}
\mathbb E_{\vec P}
\left[
\langle
A_\ell(P_\ell),
A_m(P_m)
\rangle_2
\right]
&=
\left\langle
\mathbb E_{P_\ell}A_\ell(P_\ell),
\mathbb E_{P_m}A_m(P_m)
\right\rangle_2
\nonumber\\
&=
0.
\label{eq:app_cross_terms_zero}
\end{align}
Thus all cross-layer contributions vanish in the mean squared norm.

Moreover, both Pauli conjugation and transport by \(W_{\ell-1}\) are unitary conjugations, so the Hilbert--Schmidt norm is invariant:
\begin{equation}
\|A_\ell(P_\ell)\|_2
=
\|H_\ell\|_2 .
\label{eq:app_norm_invariance}
\end{equation}
Substituting Eqs.~\eqref{eq:app_cross_terms_zero} and
\eqref{eq:app_norm_invariance} into
Eq.~\eqref{eq:app_pt_norm_expansion} yields
\begin{equation}
\mathbb E_{\vec P}
\|K(\vec P)\|_2^2
=
\sum_{\ell=1}^{L}
\|H_\ell\|_2^2 .
\label{eq:app_exact_second_moment}
\end{equation}
If the typical layerwise error strength remains \(O(1)\), then
\begin{equation}
\mathbb E_{\vec P}
\|K(\vec P)\|_2^2
=
O(L),
\label{eq:app_pt_squared_scaling}
\end{equation}
and hence
\begin{equation}
\left[
\mathbb E_{\vec P}
\|K(\vec P)\|_2^2
\right]^{1/2}
=
O(\sqrt L).
\label{eq:app_pt_sqrt_scaling}
\end{equation}

The same conclusion holds for the state-dependent action of the generator. Using Eq.~\eqref{eq:app_zero_mean} and independence between layers,
\begin{align}
\mathbb E_{\vec P}
\left\|
K(\vec P)\ket{\psi_0}
\right\|^2
&=
\mathbb E_{\vec P}
\bra{\psi_0}
K(\vec P)^2
\ket{\psi_0}
\nonumber\\
&=
\sum_{\ell=1}^{L}
\mathbb E_{P_\ell}
\bra{\psi_0}
A_\ell(P_\ell)^2
\ket{\psi_0}
\nonumber\\
&\leq
\sum_{\ell=1}^{L}
\|H_\ell\|_\infty^2
=
O(L).
\label{eq:app_state_action_scaling}
\end{align}
Therefore the root-mean-square error amplitude acting on the state is also \(O(\sqrt L)\), indicating that in the PT dressed circuit, the errors of different layers drive the state in the same directions much more slowly than the bare circuit.

Finally, Eq.~\eqref{eq:app_pt_squared_scaling} implies a simple probability bound. If
\begin{equation}
\mathbb E_{\vec P}\|K(\vec P)\|_2^2
\leq
CL
\end{equation}
for an \(L\)-independent constant \(C\), then
\begin{equation}
\Pr_{\vec P}
\left[
\|K(\vec P)\|_2
\geq
\sqrt{\frac{CL}{\delta}}
\right]
\leq
\delta
\label{eq:app_norm_markov}
\end{equation}
for any \(0<\delta<1\). Thus an independently sampled dressed realization has a characteristic quadratic-norm size \(O(\sqrt L)\), whereas a coherently aligned bare circuit can have size \(O(L)\).

\begin{figure*}[t]
\centering
\includegraphics[width=\textwidth]{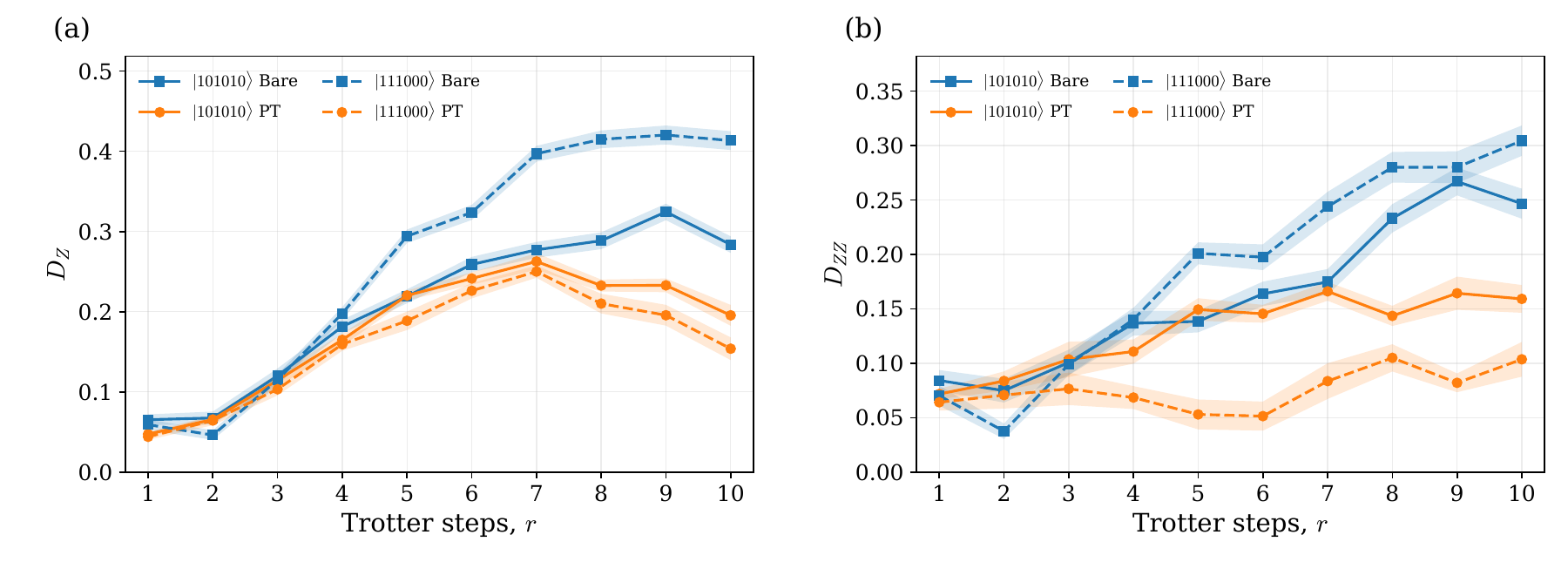}
\caption{
Profile-distance diagnostics for the Trotterized transverse-field
Ising benchmark of Sec.~\ref{subsec:hardware_tfim}, with
\(N=6\), \(J=1\), \(h=0.7\), and \(\Delta t=0.2\).
(a) Local \(Z\)-profile distance \(D_Z\), defined in
Eq.~\eqref{eq:DZ_hardware}.
(b) Nearest-neighbor \(ZZ\)-correlation distance \(D_{ZZ}\), defined
in Eq.~\eqref{eq:DZZ_hardware}.
Solid and dashed curves correspond to \(\ket{101010}\) and
\(\ket{111000}\), respectively. The PT curve shows the
mean value over many sampled realizations. Bare denotes the circuit without randomization. Shaded bands indicate $1\sigma$ standard deviation after bootstrapping the data.
}
\label{fig:tfim_hardware_profiles}
\end{figure*}

\section{Additional observables in the Trotterized Ising benchmark}
\label{app:hardware_tfim_profiles}

The imbalance benchmark in Sec.~\ref{subsec:hardware_tfim} probes
the memory of the initial spin pattern through a single weighted
sum of local magnetizations. Here we complement
Fig.~\ref{fig:tfim_hardware} by examining errors in the full local
magnetization profile and the nearest-neighbor correlation profile.
We use the same six-qubit Trotterized Ising circuits, parameters and
initial states as in the main-text benchmark.

The local \(Z\)-profile distance is
\begin{equation}
D_Z(r)
=
\left[
\frac{1}{N}
\sum_{j=0}^{N-1}
\left(
\langle Z_j\rangle_{\rm exp}
-
\langle Z_j\rangle_{\rm id}
\right)^2
\right]^{1/2},
\label{eq:DZ_hardware}
\end{equation}
and the nearest-neighbor \(ZZ\)-correlation distance is
\begin{equation}
D_{ZZ}(r)
=
\left[
\frac{1}{N-1}
\sum_{j=0}^{N-2}
\left(
\langle Z_jZ_{j+1}\rangle_{\rm exp}
-
\langle Z_jZ_{j+1}\rangle_{\rm id}
\right)^2
\right]^{1/2}.
\label{eq:DZZ_hardware}
\end{equation}
All expectation values are evaluated after \(r\) Trotter steps.
The experimental values are obtained from computational-basis
measurements, and the ideal values from the noiseless Trotter
circuit. Smaller distances indicate closer agreement with the
ideal dynamics.

Figure~\ref{fig:tfim_hardware_profiles} compares the bare circuit with PT results for both initial states. 
For the domain-wall-like input \(\ket{111000}\), the improvement is especially pronounced at larger depths. At \(r=10\), \(D_Z\) decreases from approximately \(0.413\) for the bare circuit to
\(0.154\) for PT, while \(D_{ZZ}\) decreases from approximately
\(0.304\) to \(0.104\). For the staggered input
\(\ket{101010}\), the improvement is less uniform at shallow depths but becomes evident at larger \(r\). These results complement
the imbalance measurements in  Fig.~\ref{fig:tfim_hardware} by showing that the PT benefit also extends to the mean local magnetization and nearest-neighbor correlation profiles.

\section{Quantum Fourier transform}
\label{app:hardware_qft}
\label{subsec:hardware_qft}

We also apply two-qubit-gate PT to the quantum Fourier transform \cite{shor1997polynomial,coppersmith1994approximate}. The
\(n\)-qubit QFT is defined by
\begin{equation}
U_{\rm QFT}^{(n)}
\ket{x}
=
\frac{1}{\sqrt{2^n}}
\sum_{k=0}^{2^n-1}
e^{2\pi i xk/2^n}
\ket{k},
\label{eq:qft_definition_hardware}
\end{equation}
where \(x\in\{0,1,\ldots,2^n-1\}\). Unlike the GHZ circuit, which
prepares one fixed target state, the QFT is an algorithmic unitary whose
task-level performance may depend on the chosen input. It therefore
provides a complementary test of both the generality and the
state dependence of the PT protocol.

\begin{figure*}[t]
\centering
\includegraphics[width=0.95\textwidth]{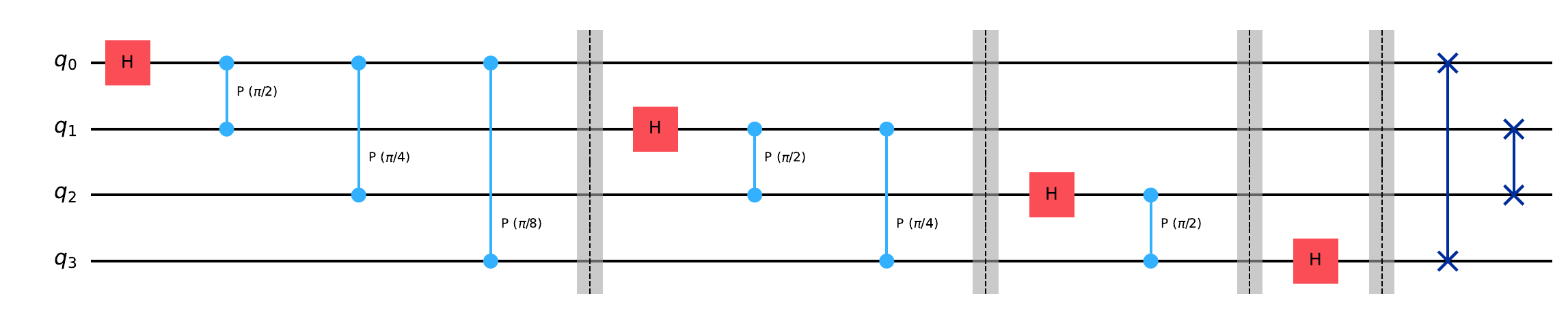}
\caption{
Four-qubit decomposed QFT circuit used in the hardware benchmark. Each
controlled-phase gate is expressed as two CNOT gates and single-qubit
phase rotations according to
Eq.~\eqref{eq:qft_cp_decomposition_hardware}. PT is
applied only to the CNOT gates, while all single-qubit phase rotations
remain unchanged.
}
\label{fig:qft_circuit}
\end{figure*}

\begin{figure}[t]
\centering
\includegraphics[width=\columnwidth]{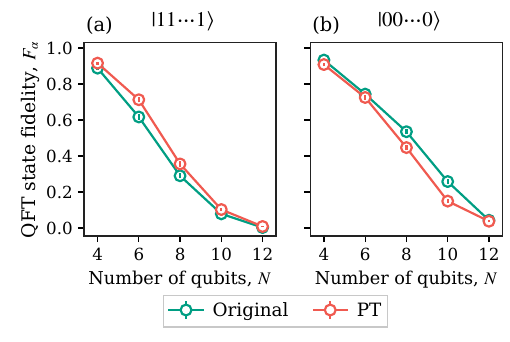}
\caption{
Input-state-dependent QFT fidelity on an IBM Pittsburgh processor
as a function of the number of qubits \(N\). Panels (a) and (b)
correspond to the computational-basis inputs
\(\ket{111\cdots1}\) and \(\ket{000\cdots0}\), respectively.
Original denotes the bare decomposed QFT circuit, and PT denotes
the mean fidelity over sampled PT realizations.
The mean improvement is evident for \(\ket{111\cdots1}\) and
absent for \(\ket{000\cdots0}\), revealing a pronounced dependence
of the task-level PT performance on the input state. Error bars show
one standard deviation estimated by bootstrapping the data.
}
\label{fig:qft_hardware}
\end{figure}

The standard QFT circuit consists of Hadamard gates, controlled-phase gates, and a final bit-reversal layer. Since a controlled-phase gate with a generic angle is non-Clifford, we first decompose it into CNOT
gates and single-qubit phase rotations. Defining
\begin{equation}
P(\phi)
=
\begin{pmatrix}
1&0\\
0&e^{i\phi}
\end{pmatrix},
\qquad
CP_{c,t}(\theta)
=
{\rm diag}(1,1,1,e^{i\theta}),
\end{equation}
we use the decomposition \cite{barenco1995elementary}
\begin{equation}
CP_{c,t}(\theta)
=
P_c\left(\frac{\theta}{2}\right)
CX_{c,t}
P_t\left(-\frac{\theta}{2}\right)
CX_{c,t}
P_t\left(\frac{\theta}{2}\right).
\label{eq:qft_cp_decomposition_hardware}
\end{equation}
The single-qubit phase gates are kept fixed, and PT is
applied independently to the two CNOT gates in each decomposed
controlled-phase block. A four-qubit example of the resulting circuit
is shown in Fig.~\ref{fig:qft_circuit}.

For a computational-basis input \(\ket{x}\), the ideal output is
\begin{equation}
\ket{\psi_x^{\rm id}}
=
U_{\rm QFT}^{(n)}\ket{x}.
\end{equation}
If the experimentally prepared output state is \(\rho_x\), we define
the input-state-dependent fidelity
\begin{equation}
F_x
=
\bra{\psi_x^{\rm id}}
\rho_x
\ket{\psi_x^{\rm id}}.
\label{eq:qft_state_fidelity_hardware}
\end{equation}
This quantity is a state-specific task fidelity rather than the process
fidelity of the full QFT channel.

For a computational-basis input, the ideal QFT output is a product of
single-qubit phase states,
\begin{equation}
\ket{\psi_x^{\rm id}}
=
\bigotimes_{j=0}^{n-1}
\ket{\varphi_j(x)},
\qquad
\ket{\varphi_j(x)}
=
\frac{
\ket{0}
+
e^{i\phi_j(x)}\ket{1}
}{\sqrt{2}},
\label{eq:qft_product_output_hardware}
\end{equation}
up to the adopted output-qubit ordering. We construct a product
verification layer
\begin{equation}
V_x
=
\bigotimes_{j=0}^{n-1}
V_{j,x},
\qquad
V_{j,x}\ket{\varphi_j(x)}
=
\ket{0}.
\label{eq:qft_verifier}
\end{equation}
For an ideal verification layer,
\begin{equation}
F_x = {\rm Tr}
\left[
V_x\rho_xV_x^\dagger
\ket{0^n}\!\bra{0^n}
\right]
\nonumber = {\rm Pr}(0^n).
\label{eq:qft_verification_probability}
\end{equation}
The probability of the all-zero outcome after the verification layer
therefore provides an experimentally accessible fidelity estimator.
On hardware, this operational estimate also includes residual errors
from the single-qubit verification gates and readout.

The measured QFT fidelities are shown in
Fig.~\ref{fig:qft_hardware}. PT denotes the mean fidelity over sampled PT realizations. For the input \(\ket{111\cdots1}\), the mean PT fidelity is systematically higher than the bare fidelity.
At \(N=8\), the bare circuit gives \(F_x=0.290\), whereas the mean
PT fidelity is \(0.356\). At \(N=10\), the corresponding values
are \(0.078\) and \(0.102\). In contrast, for \(\ket{000\cdots0}\), the PT dressing does not
improve the fidelity. At \(N=8\), the bare fidelity is
\(0.535\), whereas the mean PT fidelity is \(0.447\).

The origin of this input dependence can be understood from the gate ordering used in the QFT circuit. When a controlled-phase gate
acts, its control qubit has not yet undergone its Hadamard gate and
therefore remains in its input computational-basis state. Let
\(x_k\in\{0,1\}\) be the value of such a control qubit and let
\(\ket{\varphi_j}\) denote the state of the target qubit. The ideal
controlled-phase action is then
\begin{equation}
CP_{k,j}(\theta)
\left(
\ket{x_k}\otimes\ket{\varphi_j}
\right)
=
\ket{x_k}
\otimes
P_j(x_k\theta)
\ket{\varphi_j}.
\label{eq:qft_conditional_phase_action}
\end{equation}
If \(x_k=0\), this block acts as the identity on the ideal state at that
point in the circuit. If \(x_k=1\), it writes a nonzero phase onto the
target qubit. Consequently, every controlled-phase block has a nontrivial ideal
action for \(\ket{111\cdots1}\), while all such blocks are logically
trivial along the ideal trajectory for \(\ket{000\cdots0}\). In the last
case, the ideal QFT output reduces to
\(\ket{+}^{\otimes n}\), up to the final bit ordering.

Thus, for the all-zero input, since each ideal controlled-phase block is a logically idle operation, PT has no reason to improve the result: there is little constructive coherent error to remove, while the additional single-qubit dressing operations and the extended circuit running time may bring more error to the circuit. Another way to understand this behavior is through the  covariance matrix. Within the framework of Sec.~\ref{subsec:single_realization}, the appropriate
connected covariance is itself state dependent. For the present task, the input state \(\ket{111\cdots1}\)
exhibits larger coherent covariance, while for the input state \(\ket{000\cdots0}\), the covariance is nearly zero, which is consistent with the observed data.

The QFT experiment therefore demonstrates that two-qubit-gate PT can improve an algorithmic circuit,
while also showing that the magnitude and even the sign of the resulting performance change are determined by the state-dependent coherent-error structure probed by the chosen input. The advantage is clearly manifested only when the two-qubit gates play a nontrivial role in the ideal circuit dynamics.

\bibliographystyle{apsrev4-2}
\bibliography{ref}

\end{document}